\documentclass[preprint,12pt,3p]{elsarticle}

\usepackage{graphicx}
\graphicspath{ {figures/}{figures/pdf/} }
\usepackage{amssymb}
\usepackage{amsthm}

\usepackage{amsmath}

\usepackage{aas_macros}
\usepackage{hyperref}
\hypersetup{
  colorlinks,
  citecolor=Blue,
  linkcolor=Red,
  urlcolor=Blue}

\usepackage[nameinlink,noabbrev]{cleveref}

\usepackage{epsfig}
\usepackage{pdflscape}					

\usepackage{longtable}
\usepackage{float}
\usepackage{fancyhdr}

\fancypagestyle{mylandscape}{
	\fancyhf{} 
	\fancyfoot{
    \makebox[\textwidth][r]{
      \rlap{\hspace{.75cm}
        \smash{
          \raisebox{4.87in}{
            \rotatebox{90}{\thepage}}}}}}
}

\usepackage{cancel}

\floatstyle{plaintop}
\restylefloat{table}

\usepackage{caption}
\usepackage{subcaption}
\usepackage{afterpage}

\usepackage{amssymb}
\usepackage{amsthm}
\usepackage{amsmath}
\usepackage{xfrac}

\usepackage{color, colortbl}
\definecolor{Gray}{gray}{0.9}

\usepackage{fancyhdr} 
\usepackage[detect-all]{siunitx}

\renewcommand{\deg}{^\circ}

\newcommand{\arcsec}{^{\prime\prime}}

\newcommand{\au}{\,\mathrm{au}}
\newcommand{\km}{\,\mathrm{km}}
\newcommand{\meter}{\,\mathrm{m}}

\newcommand{\hours}{\,\mathrm{hrs}}

\newcommand{\second}{\,\mathrm{s}}

\newcommand{\K}{\,\mathrm{K}}
\newcommand{\J}{\,\mathrm{J}}

\biboptions{authoryear,round}

\journal{Planetary Science Journal}

\begin{document}

\begin{frontmatter}

\title{Thermophysical Investigation of Asteroid Surfaces I: Characterization of Thermal Inertia}



\author[label1,label2]{Eric M. MacLennan\corref{cor1}}
\ead{eric.maclennan@helsinki.fi}
\ead{mv.helsinki.fi/home/maceric/home.html}

\author[label1,label3]{Joshua P. Emery}

\cortext[cor1]{corresponding author}

\address[label1]{Earth and Planetary Sciences Department, Planetary Geosciences Institute, The University of Tennessee, Knoxville, TN 37996, USA}
\address[label2]{Department of Physics, P.O. Box 64, 00560 University of Helsinki, Finland}
\address[label3]{Department of Physics and Astronomy, Northern Arizona University, NAU Box 6010, Flagstaff, AZ 86011, USA}

\begin{abstract}

The thermal inertia of an asteroid is an indicator of the thermophysical properties of the regolith and is determined by the size of grains on the surface. Previous thermophysical modeling studies of asteroids have identified or suggested that object size, rotation period, and heliocentric distance (a proxy for temperature) are important factors that separately influence thermal inertia. In this work we present new thermal inertia values for 239 asteroids and model all three factors in a multi-variate model of thermal inertia. Using multi-epoch infrared data of this large set of objects observed by WISE, we derive the size, albedo, thermal inertia, surface roughness, and sense of spin using a thermophysical modelling approach that doesn't require {\it a priori} knowledge of an object's shape or spin axis direction. Our thermal inertia results are consistent with previous values from the literature for similarly sized asteroids, and we identify an excess of retrograde rotators among main-belt asteroids $<8\km$. We then combine our results with thermal inertias of 220 objects from the literature to construct a multi-variate model and quantify the dependency on asteroid diameter, rotation period, and surface temperature. This multi-variate model, which accounts for co-dependencies between the three independent variables, identifies asteroid diameter and surface temperature as strong controls on thermal inertia.

\end{abstract}


\end{frontmatter}


\section{Introduction}\label{S:1}

The thermophysical characterization of regolith---the unconsolidated, heterogeneous, rocky material covering the surface of other planetary bodies---is an important part of understanding the processes and evolution of airless bodies of the Solar System. By comparing thermal observations to thermophysical models, the regoliths of asteroids can be characterized by their thermal inertia ($\Gamma$). Thermal inertia is defined as $\Gamma = \sqrt{k \rho c_s}$, where $k$ is the effective thermal conductivity of the regolith, $\rho$ is the bulk density, and $c_s$ is the bulk specific heat capacity. Thermophysical models (TPMs) are often used to derive thermal inertia of a body by comparing the observed fluxes to those estimated from the model.

The largest asteroids (ranging from $\sim500-1000 \km$) in the Solar System---e.g., (1) Ceres, (2) Pallas, and (4) Vesta---exhibit global\footnote{Anomalously high thermal inertias have been measured for some localized areas on Ceres, for example.} thermal inertias \citep{Muller&Lagerros98,Capria_etal14,AliLagoa_etal20,Rognini_etal19} that are comparable to the Moon \citep[$\approx 50 \J \meter^{-2} \K^{-1} \second^{-1/2}$;][]{Wesselink48,Cremers75,Bandfield_etal11,Hayne_etal17}. On the other hand, smaller asteroids such as (433) Eros, (162173) Ryugu, (101955) Bennu, and (25143) Itokawa have approximate, estimated thermal inertias of, respectively 150, 225, 350, and $700 \J \meter^{-2} \K^{-1} \second^{-1/2}$ \citep{Mueller07,DellaGiustina_etal19,Shimaki_etal20}. The observed correlation among asteroids of thermal inertia with size \citep{Delbo_etal15,Hanus_etal18,Rozitis_etal18} suggests that only larger asteroids harbor a fine-grained regolith, whereas smaller asteroid regoliths are comprised of coarse sand-sized particles and/or a higher fraction of blocky material. This general trend suggest that asteroid size is a large factor in determining the state of asteroid regolith.

In addition to asteroid size, the rotation period has been suggested as a factor that controls asteroid thermal inertia. \cite{Harris&Drube16} used a thermal inertia estimator based on the near-Earth Asteroid Thermal Model \citep[NEATM;][]{Harris98} and found a positive correlation between rotation period, $P_\mathit{rot}$, and thermal inertia. This correlation was attributed to deeper penetration of the thermal wave into subsurface material that was higher in thermal conductivity and/or bulk density (caused by smaller porosities). The thermal wave can be expressed in terms of the thermal skin depth, $l_s$, which is the length scale over which the diurnal temperature variation changes by a factor of $e\approx2.71828$: $l_s = \sqrt{k P_\mathit{rot}/2\pi\rho c_s}$. On the other hand, \citet{Marciniak_etal19} used a TPM to derive thermal inertias of slow-rotators and found no correlation between the two variables.

For airless bodies, the effective thermal conductivity is comprised of a solid and radiative component that correspond to different heat transport mechanisms in a regolith \citep[e.g.,][]{Piqueux&Christensen09}. If radiation is the dominant form of heat propagation then thermal inertia is expected to vary as $T^{3/2}$ \citep{Delbo_etal15} because the radiative conductivity is proportional to $T^3$ \citep{Vasavada_etal99}. \citet{Rozitis_etal18} characterized the thermal inertia variation with heliocentric distance (as a proxy for temperature) for three individual asteroids and found a wide range of scaling dependencies for each. In particular, they found that the thermal inertia of two of the studied asteroids had stronger dependencies on heliocentric distance than the $T^{3/2}$ scaling law. On the other hand, the thermal inertia of Bennu as measured by the OSIRIS-REx spacecraft showed no evidence of temperature dependence \citep{Rozitis_etal20}.

Generally speaking, the temperature distribution of a surface is influenced by the thermal inertia. All other factors kept constant, higher thermal inertia surfaces have a smaller difference in sunlit and nighttime temperature, while low thermal inertia surfaces exhibit a greater diurnal temperature differences. As demonstrated in \citet{MacLennan+Emery19}, thermal emission observations taken at pre- and post-opposition (multi-epoch) are sensitive to differences in the temperature distribution and thus can be used to estimate the thermal inertia. This technique is effective at estimating the thermal inertia when no shape or spin axis information about an asteroid is known {\it a priori}.

The sense of spin (i.e. retrograde or prograde) for an asteroid can also be estimated using multi-epoch observations \citep{Mueller07,MacLennan+Emery19}. Because there exists a time-lag between the period of maximum heating at local noon and when the maximum surface temperature is reached, a morning-afternoon dichotomy is present on surfaces with a non-zero thermal inertia. Thus, the morning and afternoon sides will correspond, respectively, or inversely, to pre- and post-opposition viewing aspects, depending on the object's sense of rotation. 

In this work we use multi-epoch observations and the methods of \cite{MacLennan+Emery19} to derive thermal inertia and size estimated for 239 asteroids. In some cases, we constrain the roughness and object's sense of spin. Comparing our results to the benchmark study of \cite{MacLennan+Emery19} and other literature works, we assess the ability of this technique to estimate these TPM parameters. We then incorporate diameter, rotation period, and temperature into a unifying multi-factor thermal inertia model that simultaneously accounts for these variables. In a follow-up work, we produce grain size estimates from the thermal inertia values presented here and investigate compositional differences in regolith properties.

\section{Observations \& Thermophysical Modeling}\label{S:3}

Data from the Wide-field Infrared Survey Explorer (WISE) are used for model fitting. Absolute magnitude ($H_V$) and slope parameter ($G_V$) from \cite{Oszkiewicz_etal11}, and $P_\mathit{rot}$ from the Asteroid Lightcurve Database \citep[ALCDB;][]{Warner_etal09} are used as TPM input values for each object (\autoref{tab:HGinput}) along with mean and peak-to-trough fluxes calculated from sparse lightcurve data. The thermophysical modeling approach presented in \cite{MacLennan+Emery19} is used, as briefly summarized below, and we thus select objects that were observed by WISE at pre- and post-opposition. In \cite{MacLennan+Emery19} we extracted and used the mean and peak-to-trough flux quantities from thermal light curves via simple geometric averaging and subtracting the maximum and minimum values, respectively. Although those simplistic calculations are useful for dense lightcurve data, they can be problematic when used on sparsely sampled lightcurves for reasons discussed in \autoref{subsection2.22}.

\subsection{Data Description}\label{subsec:data}
In 2010, WISE mapped the entire sky at four photometric filters: referred to as W1, W2, W3, and W4 with wavelength centers near 3.4, 4.6, 12, and 22 $\mu$m, respectively \citep{Wright_etal10}. WISE was designed as an astrophysics all-sky mapping mission, but its infrared sensors detected the thermal emission from warm asteroids in the inner solar system. A data-processing enhancement \citep[NEOWISE;][]{Mainzer_etal11a} to the nominal pipeline was thus designed and implemented to identify and measure the emission from these solar system objects. Since the initial mapping phase in which all four bands were operational (the cryogenic phase) the WISE telescope operated at shorter wavelengths and was later reactivated (NEOWISE-R). In this work, we only use the cryogenic phase of the mission.

Since WISE does not target moving objects, the asteroids were only observed for a relatively brief (typically less than a couple days) period of time, referred to as an epoch. Each epoch of observations nominally yielded between 10 to 20 individual measurements that were separated by $\approx$1.6 hr---the orbital period of the spacecraft. NEOWISE flux data are stored at the Infrared Processing and Analysis Center\footnote{\url{http://irsa.ipac.caltech.edu/Missions/wise.html}} (IPAC) and each detection of a moving solar system object was reported to the Minor Planet Center\footnote{\url{http://www.minorplanetcenter.net/}} (MPC), where the information regarding the sky position and time of observation can be retrieved. In downloading the data, we used the MPC observation file to parse the WISE All-Sky Single Exposure (L1b) catalog on IPAC's Infrared Science Archive (IRSA) and select detections acquired within $10\second$ of that reported to the MPC, with a search cone of $10\arcsec$. We shift the isophotal wavelengths of the filters and perform a color-correction to the fluxes \citep{Wright_etal10} using a spectrum calculated from NEATM temperatures, as per the recommendation of the WISE Explanatory supplement \cite{Cutri_etal12}. Since the criteria used to parse IPAC can potentially return contaminated (i.e., by a background star or galaxy) or unwanted (non-asteroid) infrared sources from the catalog, we employ Peirce's Criterion \citep{Peirce1852,Gould1855} on the infrared color, W4--W3, as detailed in \cite{MacLennan+Emery19}, to better ensure the inclusion of only uncontaminated observations of asteroids.

\subsection{Sparse Lightcurve Sampling}\label{subsection2.22}
Due to the nature of WISE's orbit and survey cadence, a given asteroid will be observed an average of a dozen times during each epoch. This sparse sampling does not allow for the construction of a well-characterized rotational lightcurve. Since observations are taken at irregularly-spaced rotational phases---depending both on the number of observations and the object's rotation period---information may be missing for crucial points of an object's lightcurve, such as the minima and maxima. The WISE telescope orbital cadence may over-sample certain rotational phases, which poses a challenge for extracting scientifically-important characteristics such as the mean and peak-to-trough range of the lightcurve. We present here a technique for extracting these parameters from a statistically-scant photometric set, given {\it a priori} knowledge of the object's rotation period. We note that applying this approach to the objects in \citet{MacLennan+Emery19} does not significantly change the results of that work. The formulations below are similar to the analytical solution of a least-squares sinusoidal fit to lightcurve data, with some differences. We show the results of this technique on (91) Aegina in \autoref{fig2.2} and report the fluxes computed from this method for all asteroids studied in this work, along with observing circumstances, in \autoref{tab:WISEfluxes2}.

First, we step through each possible pair of flux measurement points and compute their average and difference so that for the i$th$ and j$th$ point the mean and range (absolute difference) are $m_{ij}$ and $r_{ij}$, respectively. The flux uncertainties ($\delta f$) are summed in quadrature, so that the errors in each mean ($\delta{m_{ij}}$) and range ($\delta{r_{ij}}$) are given by:

\begin{equation}\label{eq2.1}
2 \delta{m_{ij}} = \delta{r_{ij}} = \sqrt{\delta f^2_i + \delta f^2_j}
\end{equation}

Note the factor of 2 associated with the mean, as per the rules of error propagation. Proceeding, we calculate a weighting factor, $s_{ij}$, based on the separation in rotational phase (normalized to 2$\pi$ radians) of the two points, $\phi_i-\phi_j$. In this weighting scheme, pairs that sample around the same rotational phase or half a turn ($\phi_i-\phi_j = 0, 1/2$), are given a weight of $s_{ij} = 0$, and pairs separated by a quarter turn ($\phi_i-\phi_j = 1/4, 3/4$) have $s_{ij} = 1$, with linear scaling of the weights between these two extremes (top right panel of \autoref{fig2.2}).

The weighted flux mean ($\overline{F}$) and error ($\delta \overline{F}$) are then given by:
\begin{equation}\label{eq2.2}
	\overline{F} = \frac{\sum \limits_i \sum \limits_{j > i} m_{ij} \delta{m_{ij}}^{-2} s^2_{ij}}{\sum \limits_i \sum \limits_{j > i} \delta{m_{ij}}^{-2} s^2_{ij}}
\end{equation}
and
\begin{equation}\label{eq2.3}
	\delta \overline{F} = \sqrt{\ \raisebox{3pt}{$\sum \limits_{i} \sum \limits_{j > i} \delta{m_{ij}}^{2} s^2_{ij}$} \raisebox{-4pt}{\bigg/} \raisebox{-3pt}{$ \sum \limits_{i} \sum \limits_{j > i} s^2_{ij}$}}.
\end{equation}
Pair means are shown in the bottom left panel of \autoref{fig2.2}, along with the result of applying \autoref{eq2.3} to the data.

In order to formulate the lightcurve range ($\diamondsuit{F}$) and error ($\delta \diamondsuit{F}$) we employ a slightly different approach than that used for the mean. For the i$th$ point, we iterate across every combination of differences between points, to select the j$th$ point that which maximizes the range between the two: $r_{\widehat{ij}}$. Difference pairs that are separated by a quarter-turn of the asteroid are given more weight based off the pair weight, $s_{ij}$, from above (i.e., the factor $(1-\cos(4\pi s_{\widehat{ij}}))^2$):

\begin{equation}\label{eq2.4}
\diamondsuit{F} = 2\ \dfrac{\sum \limits {r_{\widehat{ij}}\delta{r_{\widehat{ij}}}^{-2}(1-\cos(4\pi s_{\widehat{ij}}))}}{\sum {\delta{r_{\widehat{ij}}}^{-2}} (1-\cos(4\pi s_{\widehat{ij}}))^2}
\end{equation}
and
\begin{equation}\label{eq2.5}
\delta \diamondsuit{F} = \sqrt{\ \raisebox{2pt}{$\sum \delta{r_{\widehat{ij}}}^{2} s^2_{\widehat{ij}}$} \raisebox{-1pt}{\Big/} \raisebox{-3pt}{$\sum s^{2}_{\widehat{ij}}$}}
\end{equation}
The factor, $1-\cos(4\pi s_{\widehat{ij}})$, is used to scale the $s_{\widehat{ij}}$ factor in \autoref{eq2.4} in order to create a weight function based off a sinusoid, as opposed to a linear relationship, because we wish to add weights to the error estimation that are appropriate for a rotating non-spherical object. However, this factor is not used in \autoref{eq2.5} since doing so would create a penalty for data that do not resemble a sine function, such as for shapes that significantly deviate from an ellipsoid. Employing \autoref{eq2.4} is essentially the same as extracting the peak-to-trough range of the the best-fit sinusoid to the original lightcurve. Pair ranges are shown in the bottom right panel of \autoref{fig2.2} with the function given in \autoref{eq2.5} in red.

\subsection{TPM Implementation}

The TPM and data-fitting approach used here is identical to that presented in \cite{MacLennan+Emery19} and is summarized briefly here. First, the surface temperatures are modeled across the surface of a spherical object constructed of discrete facets. The one-dimensional heat transfer equation (Fourier's Law) is numerically solved  using the estimated insolation (incoming solar radiation) as the energy input. The discrete facets are characterized as planar faces and divided into latitude bins. A diurnal cycle is simulated by rotating the facets about the object's spin axis. Two types of surfaces are modeled: a perfectly smooth surface in which only direct insolation is considered, and a rough surface that is comprised of spherical-section craters, for which direct and multiply-scattered insolation and thermally re-radiated energy from other facets are calculated. Surface roughness is characterized by the mean surface slope \citep[$\bar{\theta}$;][]{Hapke84}, which is varied by differing both the opening angle of the crater ($\gamma$) and the proportion of surface area that is covered by those craters ($f_R$); the latter is implemented when calculating the flux contribution of rough and smooth surfaces.

We use parameterized forms of the energy balance equation and heat diffusion equation \citep[see][for further details]{MacLennan+Emery19}, which reduces the number of TPM variables that are necessary to calculate a unique surface temperature distribution, in order to construct temperature reference tables and reduce the computational time. In this scheme, the necessary information required for rough surface temperature calculation is the bond albedo ($A_b$), thermal parameter,
\begin{equation}\label{eq:theta}
    \Theta = \frac{\Gamma}{\varepsilon_B \sigma_0 T_{eq}^3} \sqrt{\frac{2 \pi}{P_\mathit{rot}}},
\end{equation}
and sub-solar latitude; whereas the smooth surface only requires the thermal parameter and sub-solar latitude. In \autoref{eq:theta}, $\sigma_0$ is the Stefan-Boltzmann constant, $\varepsilon_B$ is the bolometric emissivity, and $T_{eq}$ is the sub-solar equilibrium temperature:
\begin{equation}\label{eq:eqT}
    \varepsilon_B \sigma_0 T_\textrm{eq}^4 = \frac{S_\odot (1-A)}{R^2_{\au}}.
\end{equation}
In the case of smooth surfaces $A_b$ is implicitly accounted for in the $T_{eq}$ term and we thus do not need to specify it to run the TPM. In the case of rough surfaces, $A_b$ explicitly determines the amount of multiple-scattering within a crater. Thus we run the rough surface for various values of $A_b$, as detailed in the next paragraph. The surface temperatures for both the smooth and rough surface TPMs are stored in reference tables, expressed as $T' = T/T_\textrm{eq}$.

The smooth-surface TPM was run for 46 values of sub-solar latitude ($0^{\circ}$ to $90^{\circ}$ in $2^{\circ}$ increments) and 116 values of the thermal parameter (spaced equally in log$_{10}$ space, from 0 to 450) whereas the rough-surface TPM was iterated across 3 values of $\gamma =\{ 45^{\circ}, 68^{\circ}, 90^{\circ}\}$ and run for 46 values of sub-solar latitude ($0^{\circ}$ to $90^{\circ}$ in $2^{\circ}$ increments), 116 values of the thermal parameter (spread out in log$_{10}$ space, from 0 to 450), and 7 values of $A_\mathit{grid} = \{0, 0.1, 0.2, 0.3, 0.4, 0.5, 1\}$. These parameters are chosen to ensure an accuracy within 1\% between the surface temperature values interpolated from the grid and those calculated using the exact model parameters.

Surface temperatures calculated for spheres are mapped to prolate ellipsoids (\textsf{b/c} = 1, where $\textsf{a} \geq \textsf{b} \geq \textsf{c}$) using closed-form algebraic expressions \citep[i.e., Appendix B in][]{MacLennan+Emery19} in order to model elongated bodies of differing \textsf{a/b} axis ratio. Fluxes are calculated for the given observing circumstances by interpolation of the flux calculated using the tabulated temperatures. The flux calculated from the interpolated grid are within 1\% of the flux calculated by running the TPM with the exact thermophysical and observing parameters. Finally, thermal flux is calculated by a summation of the individual flux contributions from smooth surface ($B_\mathit{smooth}$) and crater elements ($B_\mathit{rough}$) and using a grey-body approximation, with spectral emissivity ($\varepsilon_{\lambda} = 0.9$):
\begin{equation}\label{eq:fluxcalc}
	F(\lambda) = \frac{\varepsilon_{\lambda}}{\Delta_{\au}^{2}} \sum\limits_{S} \{(1-f_R)B_\mathit{smooth}(\lambda, T)\cos(e) + f_R a_R v \Lambda B_\mathit{rough}(\lambda, T)\cos(e_R)\} a_f
\end{equation}
$e_\angle$ and $e_{R\angle}$ are the emission angle of the flat facet and crater element. The facet area is $a_f$, and the crater element areas are $a_R = 2/(m(1+\cos{\gamma}))$ where $m=40$ is the number of crater elements \citep{Emery_etal14}. We note that the flux calculation formula stated in \citet{MacLennan+Emery19} neglects the latter two parameters for a rough surface and that the one presented here is correct. For rough (crater) elements, $v$ is used to indicate if it is visible ($v = 1$) or not ($v = 0$) to the observer, and $\Lambda$ is a correction factor used to adjust fluxes that deviate from the pre-computed $A_\mathit{grid}$ values \citep[for more details, see][]{MacLennan+Emery19}.

In our data-fitting approach, the shape, spin vector ($\lambda_{eclip}$, $\beta_{eclip}$), roughness, and thermal inertia are left as free parameters that we select from a pre-defined sample space. A sphere and prolate ellipsoids with \textsf{a/b} axis ratios of 1.25, 1.75, 2.5, and 3.5 are used. For each of these shapes, we sample 25 predefined thermal inertia values, 3 default roughness (mean surface slope; $\bar{\theta}$) values, and 235 spin vectors. We search for the best-fit $D_\mathit{eff}$ for each combination of these parameters. Each individual value of $\gamma$ is paired with a value of $f_R = \{1/2, 4/5, 1\}$ that correspond to default mean surface slopes of $\bar{\theta} = \{10^{\circ}, 29^{\circ},$ and $58^{\circ}\}$. The thermal inertia points are uniformly distributed in log$_{10}$ space from 0 to 3000 J m$^{-2}$ K$^{-1}$ s$^{-1/2}$, and the spin vectors are spread evenly throughout the celestial sphere, which is achieved by constructing a Fibonacci lattice in spherical coordinates \citep[e.g.,][]{Swinbank&Purser06}. For each shape/spin vector/$\bar{\theta}$/$\Gamma$ combination we use a routine to find the $D_\mathit{eff}$ value which minimizes $\chi^{2}$. To place confidence limits on each of the fitted parameters, we use the reduced $\chi^2$ statistic ($\tilde{\chi}^2 = \sfrac{\chi^2}{\nu}$) to express the solutions within a 1$\sigma$ range as $\tilde{\chi}^2 < \tilde{\chi}^2_\mathit{min}(1+\sfrac{\sqrt{2\nu}}{\nu})$ and consider solutions with $\tilde{\chi}^2_\mathit{min} < 8$ to be acceptable.

\subsection{Characteristic Temperature Calculation}\label{colorT}

Estimating the temperature of an asteroid at the time of observation is necessary to perform our multivariate analysis. Calculating a single value for the characteristic surface temperature of an asteroid, which exhibits wide temperature variations across the surface, can be approached in a few different ways. One approach is to rely on the estimation of the sub-solar temperature, based on the theoretical energy balance formulation (\autoref{eq:eqT}). This approach has two problems: the assumptions made in the energy balance equation will often lead to the  overestimation of the true sub-solar temperature, and there is a low likelihood that the sub-observer point is close to the sub-solar point.

In order to overcome these possible problems we calculate the color temperature, $T_c$, by independently fitting a blackbody curve (via least-squares minimization) to the asteroid's W3 and W4 thermal fluxes. This approach implicitly accounts for the spatial variation in surface temperatures and explicitly calculated from the data itself, as opposed to using the sub-solar temperature, $T_{ss}$. We found that a blackbody assumption ($\varepsilon_B = 1$) does not introduce uncertainty in the temperature, as any non-zero value would not shift the peak of a blackbody emission curve, which is related to the temperature through Wein's Law. Fitting a blackbody function directly to the WISE dataset is straightforward, but retroactively applying this approach to thermal inertias found in the literature is less so. To estimate $T_c$ for asteroids in the literature we calculate the relationship between the NEATM sub-solar temperature $T_{ss}$\footnote{This was calculated during the calibration of WISE data for each observation and then averaged for each object.} (\autoref{eq:NEATMbb}) and $T_c$ for our set of asteroids: $T_{ss} = 0.777 \times T_{c}^{1.063 (\pm 0.005)}$. This best-fit equation, along with the data, is depicted in \autoref{figTcolorneatm} by the dotted red line and blue dot-dash lines showing the 1$\sigma$ uncertainty bounds in the exponent. We invert this equation and use it on the TPM results from previous works, since it is often not the case that a temperature is reported with the thermal inertia. For these objects we use,
\begin{equation}\label{eq:NEATMbb}
	T_{ss} = \bigg[ \frac{S_{\odot}(1 - A)}{\eta \varepsilon_B \sigma_0 R_{\au}^2} \bigg]^{1/4},
\end{equation}
and assume a beaming parameter of $\eta = 1.1$ \citep[the approximate mean for main-belt objects][]{Mainzer_etal11b}, to compute $T_{ss}$ and then $T_c$.

\section{Results and Analysis}\label{sec:TPMresults}

The TPM was run for 239 objects: 3 near-Earth asteroids (NEAs), 2 Mars-crossers (MCs), and 234 main-belt asteorids (MBAs). \autoref{tab:TPMresults} shows the best-fit and 1$\sigma$ uncertainties for the effective diameter ($D_\mathit{eff}$), geometric albedo ($p_V$), thermal inertia ($\Gamma$), surface roughness ($\bar{\theta}$), elongation (\textsf{a/b}; prolate ellipsoid axis ratio), and sense of spin ($\Uparrow$ for prograde and $\Downarrow$ for retrograde) for all 239 objects, with the results of the 21 object from \cite{MacLennan+Emery19} included at the top. Diameter errors for $D_\mathit{eff} > 10$ km are below 15\% of the diameter value, but can be as high as 40\% for objects smaller than 10 km. Upper and lower thermal inertia uncertainties are, on average, 180\% and 67\% of the reported value, respectively. Surface roughness could only be estimated for 97 of the 239 (41\%) objects. Informed by the model validation tests of \citet{MacLennan+Emery19}, we use spherical shapes to make an estimate of the sense of spin, which could be unambiguously estimated for all but 17 of the 239 (93\%) objects. In some cases, TPM fits only allowed for a lower or upper bound on the surface roughness. Note that objects with TPM fits having $\tilde{\chi}^2_\mathit{min} > 8$ are marked in \autoref{tab:TPMresults} and should be used with caution.

We combine our sense of spin results with that in \citet{MacLennan+Emery19} and compare to the spin poles of object shapes that are in the DAMIT\footnote{http://astro.troja.mff.cuni.cz/projects/asteroids3D/web.php} database \citep[Database of Asteroid Models from Inversion Techniques;][]{Durech10}. In total, there are 101 objects both datasets and 77 of them have sense of spin estimate that agree. If we assume that the DAMIT spin axis estimates have 100\% accurate sense of spin, then the TPM has a $76.2\% \pm 4.3\%$ success rate, based on binomial probability distribution. \citet{MacLennan+Emery19} demonstrated that the sense of spin success rate is dependent on the thermal inertia, with a success rate of $65-80\%$ in the range $\Gamma = 40-150 \J \meter^{-2} \K^{-1} \second^{-1/2}$ when using spheres. The fact that this agrees with our comparison in this work is encouraging, yet more investigation into model development should be performed in an effort to improve the success rate of constraining the sense of spin using TPMs.

Based on the TPM results, we observe a correlation between the retrograde/prograde ratio and asteroid size. We bin our set of objects by diameter in \autoref{fig:proretro} and assign the uncertainty (shown as vertical lines) of the bins to be the number of objects with indeterminate spins in that diameter bin, or the $\sim76\%$ success rate based on our check with DAMIT spins---whichever is larger. The horizontal lines that transect some of the spin/diameter bins indicate the number of NEAs in that bin. The fraction of prograde to retrograde rotators in most of the bins in our sample are statistically-indistinguishable. Only asteroids with $D_\mathit{eff} < 8 \km$ show a statistically-significant excess of retrograde rotators, which we discuss in \autoref{sec:discussion}.

The diameter estimates from NEATM fits presented by the WISE team \citep[i.e.,][]{Mainzer_etal11b,Masiero_etal11} are reported for each sighting, or epoch, which can have pre- or post-opposition geometry. Because the NEATM assumes a spherical shape, it is most-useful to compare the volume-equivalent, effective diameters of the ellipsoid to their values. We present a comparison of these diameter pairs ($D^\textrm{WISE}_\mathit{eff}$) to the diameter values (one per object; $D^\textrm{TPM}_\mathit{eff}$) obtained here, and plot them (colored by observing geometry) in \autoref{fig:diamcomp2}. There is a general agreement of within $\pm$ 15\% between the two datasets, with a few important notes. Firstly, for objects $\sim$30 km and above, our TPM diameters are slightly higher than the NEATM model estimates of the WISE team. This discrepancy is likely due to the inherent model differences between our TPM approach and the NEATM used by the WISE team. Secondly, objects smaller than 20 km exhibit, on average, 5\% lower diameters from our TPM analysis than from the WISE NEATM analysis. Lastly, we highlight an interesting trend seen in \autoref{fig:diamcomp2} for different observing geometries: pre-opposition (upright triangles) NEATM diameters are more similar to the TPM-derived diameters for objects smaller than $\sim8~\km$ while the post-opposition (downward triangles) diameters remain consistently offset from our TPM diameters at smaller sizes. From this result we can conclude that the majority (over 50\%) of small diameter asteroids are retrograde rotators---which serves as an independent check on our sense of spin results (\autoref{sec:TPMresults}). 

A handful of asteroids with thermal inertias presented here have previous estimates from the works of \citet{Hanus_etal18}, \citet{Marciniak_etal19}, and \citet{Pravec_etal19}. We depict all these estimates in \autoref{fig:TIcomp}. In several instances, two thermal inertias were reported because two shape/spin solutions were used, for which we show both values. In nearly all cases there is good agreement between our estimate and the previously-reported value (i.e., the error bars overlap). Only (1741) Giclas shows a significant difference between our estimate and the previous estimates \citep{Pravec_etal19}---our estimate is smaller by around a factor of three and there is no overlap at the 1$\sigma$ level. We note that our retrograde sense of spin estimate for Giclas is opposite to that of the prograde shape solution in \citet{Pravec_etal19}, and our roughness estimate is much higher ($\approx 58\deg$ compared to $38.8\deg$). If we were to only use the prograde solutions from out fitting, our estimate would not change, but if we consider higher roughness values that have a higher $\tilde{\chi}^2_\mathit{min}$ then the uncertainty in our estimate would overlap with \citet{Pravec_etal19}. This difference in thermal inertia may most likely be caused by the ellipsoidal shape assumption used in TPM fitting.

In addition to comparing the diameters and themal inertias of individual objects from our dataset to the findings from the WISE team, we compare thermal inertia results of 220 asteroids from previous TPM works. Combined with the results from \cite{MacLennan+Emery19}, we present thermal inertia estimates for 250 asteroids (19 of which have previous determinations in other works), an approximate doubling over the tally of literature values (\autoref{tab:prevTPM})---mostly in the 5--$50\km$ size range. We highlight previous authors and works that have presented thermal inertia estimates for 5 objects or more, notably \cite{AliLagoa_etal20,Hanus_etal15,Hanus_etal16,Hanus_etal18,Marciniak_etal18,Marciniak_etal19}. \citet{AliLagoa_etal20} targets some of the largest asteroids in the Main-Belt. Similar to this work and \citet{MacLennan+Emery19}, \citet{Hanus_etal15} and \citet{Hanus_etal18} have collectively modeled dozens of asteroids that were observed by WISE. \cite{Marciniak_etal18} and \cite{Marciniak_etal19} specifically targeted asteroids with longer rotation periods ($P_\mathit{rot} > 24 \hours$); a group of objects that have lacked thermal inertia estimations. We refer to object thermal inertias presented in papers with less than 5 objects as ``miscellaneous literature''.

\subsection{Multivariate Regression Model}\label{subsec:mrv}

We implement a forward stepwise multivariate regression model \citep{Draper&Smith98} on the thermal inertias presented here and in previous works (\autoref{tab:TPMresults} and \autoref{tab:prevTPM}) in order to characterize the controlling factors. The independent factors in this model are color temperature ($T_c$)---an approximation of the surface temperature; \autoref{colorT}---object diameter ($D_\mathit{eff}$) and rotation period ($P_\mathit{rot}$). All variables, including $\Gamma$ are transformed into log$_{10}$ space when included in the model. We use the inverse of the uncertainty in $\Gamma$ as a weighting factor for each object in the model. The forward stepwise regression algorithm permits a factor to enter the model when the relationship with the dependent variable is statistically-significant ($p<.05$).

The regression model selected all input variables as statistically-significant explanatory variables. The equation is given by,
\begin{equation}\label{eq:TIdep}
    \log_{10}\Gamma = W + X(\log_{10}T_c) + Y(\log_{10}D_\mathit{eff}) + Z(\log_{10}P_\mathit{rot}),
\end{equation}
with best-fit intercept and coefficient values: $W = -4.65 \pm 0.70$, $X = 2.74 \pm 0.29$, $Y = -0.17 \pm 0.03$, and $Z = 0.12 \pm 0.05$. This best-fit model and data are shown in \autoref{fig:TImod} Previous studies quantified the thermal inertia dependence on diameter \citep[e.g.][]{Delbo&Tanga09}, rotation period \citep{Harris&Drube16}, and heliocentric distance \citep[as a proxy for temperature;][]{Rozitis_etal18} seperately, but no study to-date has attempted to {\it simultaneously} account for the effect of all three of these variables on thermal inertia. Performing a multivariate regression, as we have done here, accounts for confounding effects between variables, such as the codependency between diameter and surface temperature. This is particularly important because the smallest objects tend to be observed at smaller heliocentric distances and thus have larger surface temperatures.

\subsection{Noteworthy Objects}

Notably, two objects in this study have higher estimated thermal inertias than any other asteroid to-date:

\paragraph{(3554) Amun} Discovered in 1986, this Venus-crossing, Aten NEA has an estimated size of $D_\mathit{eff} = 2.71 \pm 0.02$ km. Its rotation period of $2.53$ hr places it close to the theoretical spin barrier limit, and near-infrared reflectance observations show a red and featureless spectrum yielding an ambiguous classification as a X- or D-type \citep{Thomas_etal14}. Our moderate albedo ($p_V \approx 0.241$) estimate and its very high thermal inertia are highly suggestive of a metal-rich surface, which may help explain the thermal inertia of $\sim1400 \J \meter^{-2} \K^{-1} \second^{-1/2}$ estimated here. Additionally, our low roughness estimate is interesting to note, as it suggests a surface that is relatively smooth at the cm scale (i.e., on the order of $l_s$).

\paragraph{(5604) 1992 FE} This V-type NEA is also an Aten and has been flagged as a Potentially Hazardous Asteroid (PHA) by the MPC. This sub-kilometer object has a very high thermal inertia, but with large error bars: $1100^{+2200}_{-600} \J \meter^{-2} \K^{-1} \second^{-1/2}$. Its high optical albedo and radar circular polarization ratio\footnote{https://echo.jpl.nasa.gov/asteroids/1992FE/1992FE\_planning.2017.html} are consistent with its V-type taxonomic classification and having surface properties similar to Vesta \citep{Benner_etal08}.

\section{Discussion}\label{sec:discussion}

Our results show a slight excess of prograde spins at larger sizes (\autoref{fig:proretro}), which is generally consistent with previous findings of spin vector distributions estimated from lightcurve inversion methods \citep{Kryszczynska_etal07,Hanus_etal11,Durech_etal16}. The estimated success rate of our sense of spin determinations places some uncertainty on this claim, however. We can be most confident about an prograde/retrograde difference in the $>60 \km$ size bin, which is consistent with the findings of \citet{Hanus_etal11,Durech_etal16}. But the large, overlapping uncertainties in the 16--$30\km$ and 30--$60 \km$ places doubt on any claim of excess prograde rotators. The prograde excess for these large objects is likely a remnant of the primordial spins of large protoplanets due to the accretion direction of pebbles into planetesimals \citep{Johansen&Lacerda10}.

Our results show a curious overabundance of small ($<8\km$) retrograde rotators. An overabundance of retrograde rotators among NEAs was presented by \cite{LaSpina_etal04}, with the cause attributed to a dynamical selection effect: retrograde MBAs are more likely to feed into resonances, via the Yarkovsky effect, that alter their orbits into near-Earth space \citep{Bottke_etal02}. Properly investigating and explaining this result is beyond the scope of this work, but we suspect that modeling of YORP spin obliquity evolution \citep[e.g.,][]{Vokrouhlicky_etal03} and/or the spin alteration due to collisions \citep{Sevecek_etal19} should be used to investigate this topic. Yet, the MBAs studied here have not yet been subjected to this dynamical selection effect. We also note that, because we only consider asteroids with rotation periods in the ALCDB, our object set is subject to the observational biases inherent in the determination of rotation periods. This includes, but is not limited to, the skew of known rotation periods to less than Earth's rotation period and object shapes that depart from spherical shapes.

The multivariate model of asteroid thermal inertia indicates that temperature is a strong controlling factor ($p < .001$). The best-fit coefficient in \autoref{eq:TIdep} can be written as the proportionality: $\Gamma \propto T^{2.74 \pm 0.29}_c$. Because surface temperatures generally scale with the inverse square of heliocentric distance (\autoref{eq:eqT}), it follows that $\Gamma \propto r^{\alpha}$. Our results can be expressed in terms of this proportionality by using $T_c^X = R_{\au}^{-2\alpha}$, which gives: $\alpha = -1.37 \pm 0.14$. If only the radiative component of thermal conductivity on thermal inertia is considered, the expected coefficient would be $\alpha = -0.75$. \cite{Rozitis_etal18} calculated $\alpha$ for three objects ranging from $-2.5$ to $-1$, with each object having a different best-fit $\alpha$. Our result is remarkably consistent with all three objects \citep[see Fig. 8 in][]{Rozitis_etal18}, although the asteroids studied in that work exhibited vastly different thermal inertia dependence on heliocentric distance. Similar to \citet{Harris&Drube16}, \citet{Rozitis_etal18} suggested that increased solar heating would allow the thermal wave to sample higher thermal inertia material in the sub-surface due to an increase in $l_s$. This would result in an increase in thermal inertia for warmer objects if the thermal conductivity and/or bulk density increases with depth---as is the case for the Moon \citep{Keihm&Langseth73}.


In the 200--$350\K$ range---the approximate range for most asteroids---the heat capacity is also temperature-dependent \citep[$c_s \propto T$;][]{Opeil_etal12,Macke_etal19} and should also contribute to the thermal inertia temperature-dependence. The overall dependence of thermal inertia on temperature should be stronger than that predicted using only the radiative component of thermal conductivity, namely $\Gamma \propto T^{2}$. When combining the temperature-dependence of the radiative component of thermal conductivity and heat capacity together into the thermal inertia dependency of temperature, the expected relationship is still weaker than the observed dependence presented here.

Our multivariate regression model for thermal inertia also selected the diameter ($p < .001$) and rotation period ($p = .011$) as statistically significant factors. The trend of increasing thermal inertia with smaller asteroid size was established by \cite{Delbo_etal07} and has been supported by subsequent works that increase the overall number of thermal inertia estimates \citep{Delbo_etal15}. \cite{Delbo&Tanga09} found a power-law exponent of $-0.21 \pm 0.4$ between diameter and thermal inertia for NEAs and MBAs with sizes $<100\km$. The \citet{Delbo&Tanga09} value is consistent with, but somewhat smaller than, our value of $-0.17 \pm 0.03$, implying a stronger relationship. It is not unexpected that our estimate is larger because we account for the temperature-dependency. For example, \cite{Rozitis_etal18} found that the diameter and temperature power-law exponents are inversely correlated, and that $\alpha = -1.32$ corresponds to a power-law diameter exponent of $-0.18$, which is consistent with our estimate of $-0.17 \pm 0.03$.

Whereas the dependency on diameter is statistically robust, our findings show that the relationship between thermal inertia and rotation period is barely distinguishable from a slope of zero ($Z = 0.12 \pm 0.05$). In the work of \cite{Harris&Drube16}, who used a NEATM-based thermal inertia estimator, found a significant correlation between thermal inertia and rotation period for asteroids with rotation periods spanning 2--$200\hours$. However, the works of \cite{Marciniak_etal18,Marciniak_etal19} found an abundance of low-$\Gamma$ slow-rotators ($P_\mathit{rot}>12\hours$) using a TPM that explicitly accounts for thermal inertia. Considering the results in this paper and from these previous works we claim that the relationship between thermal inertia and rotation period, if present, is very weak.

Future thermophysical modeling efforts should target more slow rotators to better characterize their thermal inertia and understand the its relationship (or lack thereof) with asteroid rotation period. Higher thermal inertias could be indicative of the increase in thermal conductivity (or bulk density due to compression) for objects with large $l_s$ values. These objects with large $l_s$, which include asteroids with high surface temperatures \citep{Rozitis_etal18}, can be used to investigate possible changes in regolith properties as a function of depth \citep[i.e.,][]{Harris&Drube16}.

\section{Conclusions and Follow-Up Work}\label{conclusion}

In this work, we applied the method of \citet{MacLennan+Emery19} to WISE multi-epoch observations in order to estimate the effective diameter, geometric albedo, thermal inertia, and surface roughness for 239 asteroids (\autoref{tab:TPMresults}). Additionally, we report the shape and sense of spin for a large fraction of these objects \autoref{sec:TPMresults}. Our thermal inertia estimates are consistent with previous literature values for individual objects (\autoref{fig:TIcomp}) and for objects with similar size and rotation period. From our results, we conclude that surface temperature, asteroid size (inverse relationship), and rotation period are controls of thermal inertia of asteroids. We find that the relationship between thermal inertia and size is present, but less pronounced than suggested in previous works that do not also consider the influence of temperature (\autoref{subsec:mrv}). The temperature dependence ($\Gamma \propto T^{2.74 \pm 0.29}_c$) is larger than the theoretical prediction of $\Gamma \propto T^{1.5}$ if only the temperature-dependence of the radiative component of thermal conductivity is considered, and of $\Gamma \propto T^{2}$ if the temperature-dependence of heat capacity is additionally considered. Instead, this relationship between thermal inertia and temperature is consistent with temperature-dependency of both the heat capacity and thermal conductivity (\autoref{sec:discussion}). The thermal inertia dependence on object rotation period is weak and increased statistics of slow-rotator thermal inertias in the future could either support or negate this finding.

In a follow-up work, we will utilize a thermal conductivity model to estimate characteristic grain sizes for each object in this thermal inertia dataset. These grain sizes will then be used to investigate plausible regolith development mechanisms such as impact erosion and thermal fatigue cycling. We will then run the grain sizes through a multi-variate regression model similar to that performed here in order to explore the controlling factors of regolith evolution on asteroids.

\section*{Acknowledgements}

We thank two anonymous reviewers for their thoughtful critiques which improved the presentation of this paper. E.M.M. is supported by the NASA Earth and Space Science Fellowship \#NNX14AP21H.





\bibliographystyle{apalike}
\bibliography{references-mechwxng}

\begin{landscape}
\thispagestyle{mylandscape}
\begin{figure}
  \centering
  \setlength{\tabcolsep}{2pt}
  \begin{tabular}{rl}
  		\includegraphics[clip,trim = 0.6cm 0.25cm .8cm 1cm,width=.35\linewidth]{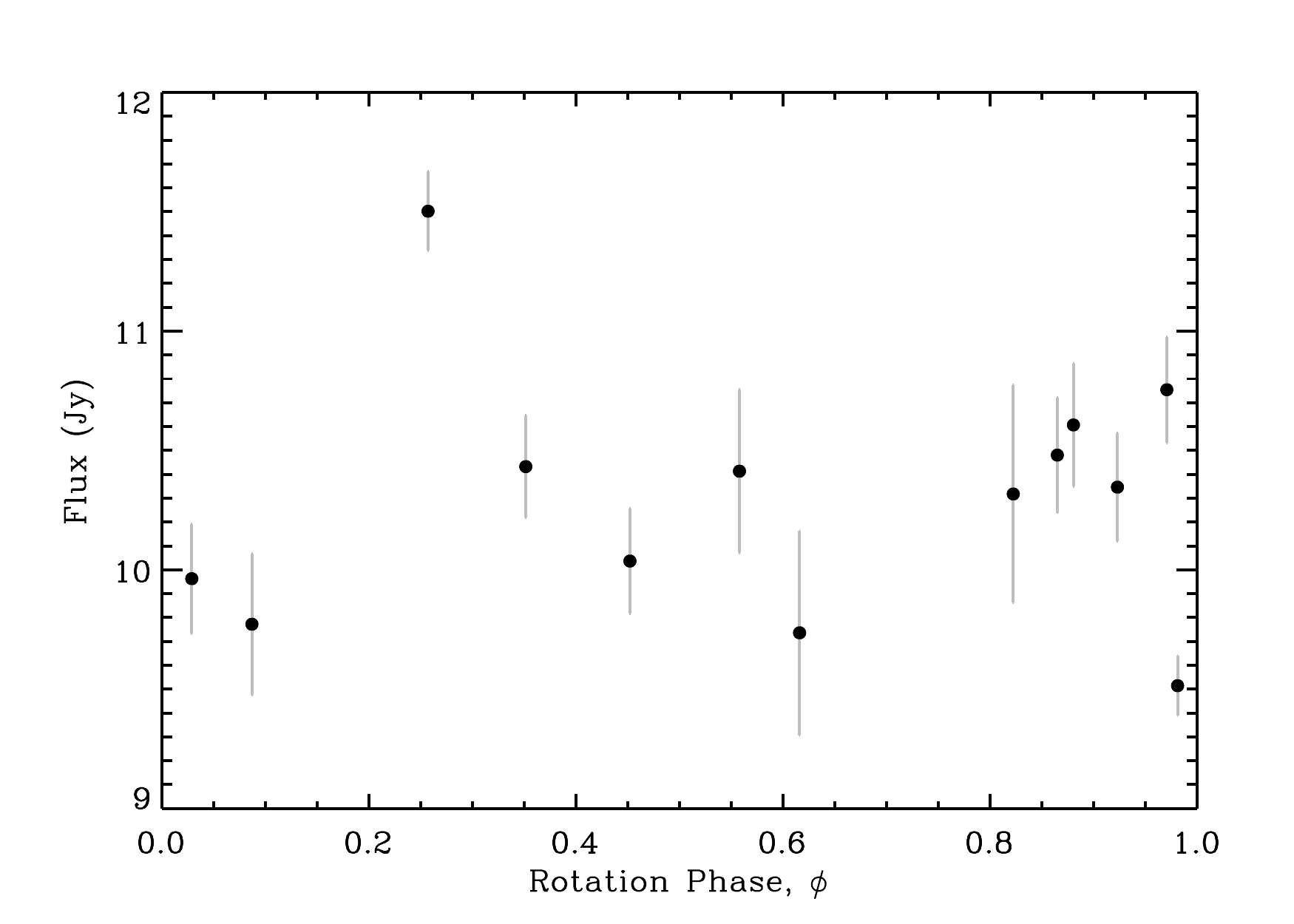} &
		\includegraphics[clip,trim = 0.6cm 0.25cm .8cm 1cm,width=.35\linewidth]{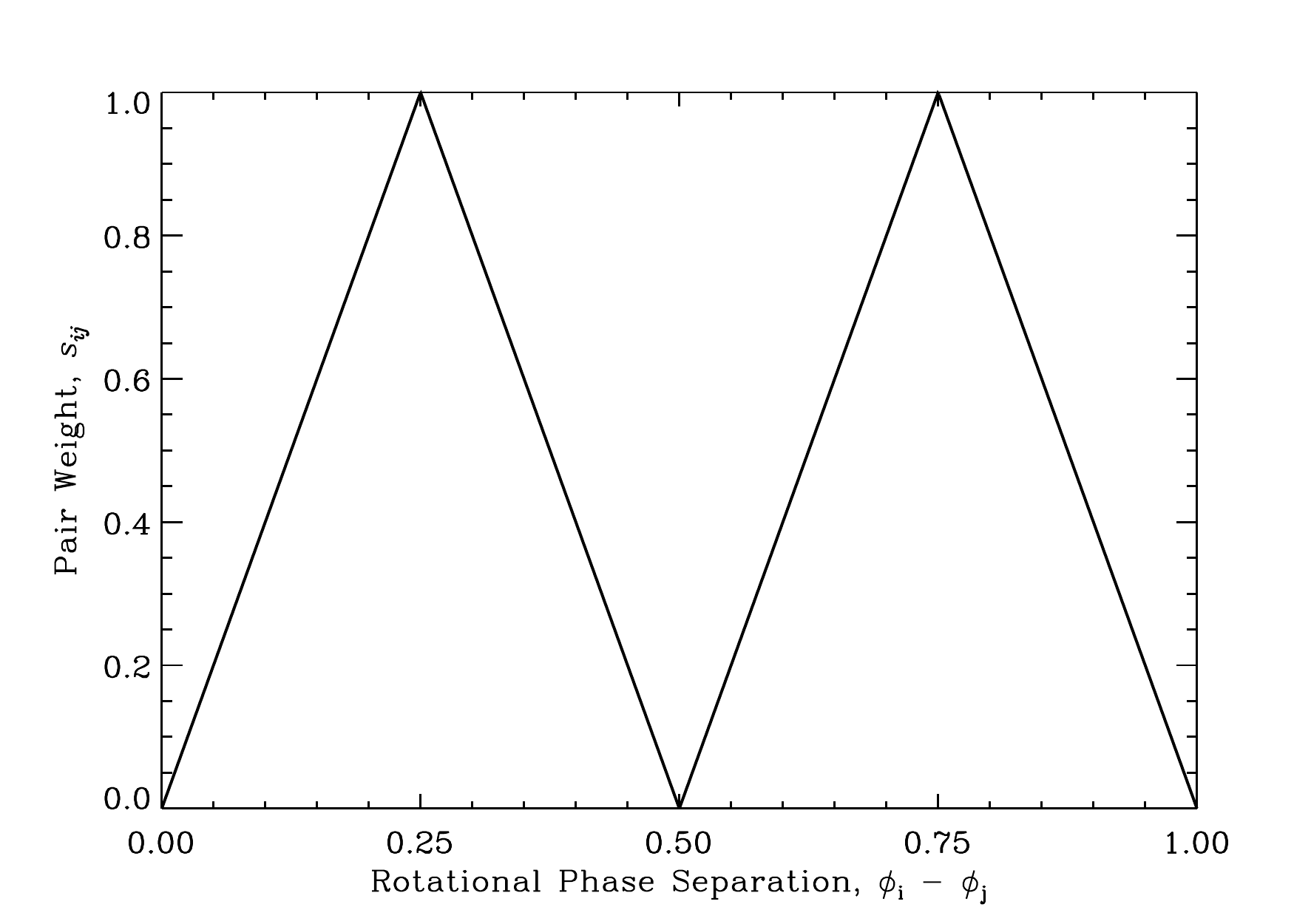} \\
		\includegraphics[clip,trim = 0.6cm 0.25cm .8cm 1cm,width=.35\linewidth]{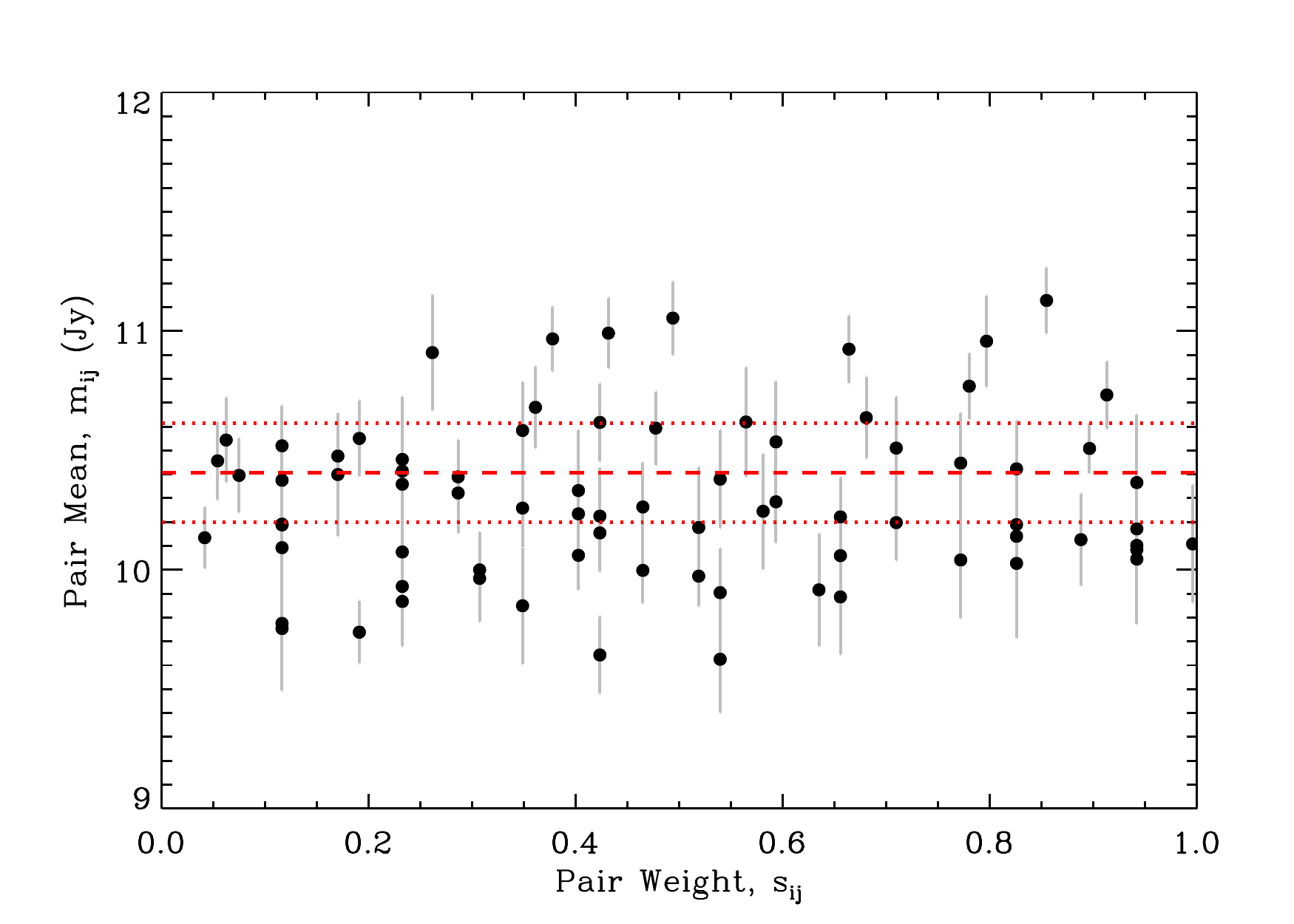} &
		\includegraphics[clip,trim = 0.6cm 0.25cm .8cm 1cm,width=.35\linewidth]{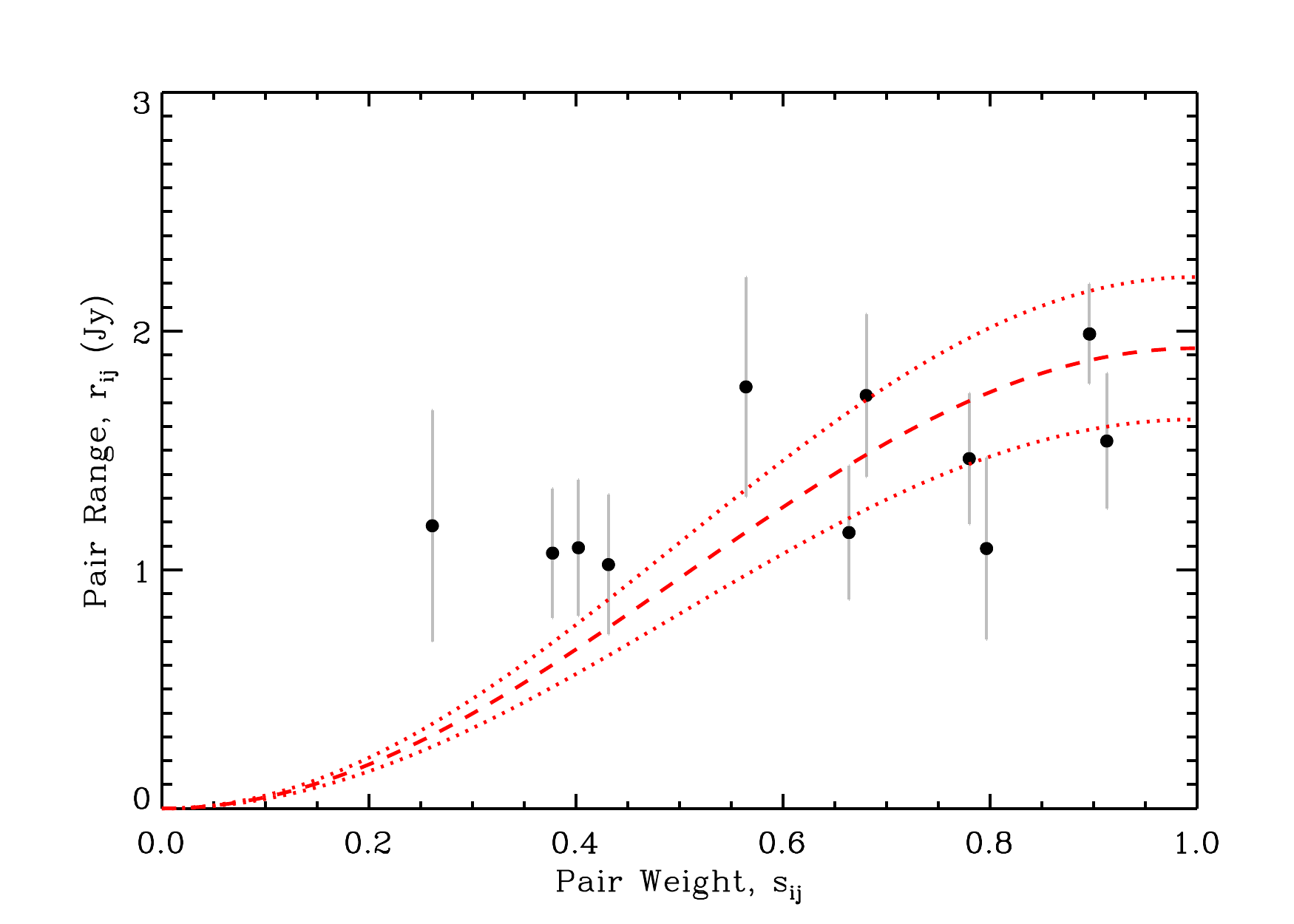}
	\end{tabular}
	\caption{Graphical depiction of applying \autoref{eq2.2} and \autoref{eq2.4} toward computing the mean (lower left) and flux range (lower right) of W4 data for (91) Aegina. The upper left panel shows the W4 fluxes as a function of rotation phase and the upper right panel depicts the weights applied to pairs of fluxes as a function of their phase separation. The red dashed line and dotted lines give the best-fit and 1$\sigma$ uncertainty for the mean and amplitude of $\overline{F} = 10.4 \pm 2.2 $ Jy and $\diamondsuit{F} = 1.9 \pm 0.3$ Jy.}\label{fig2.2}
\end{figure}
\end{landscape}

\begin{figure}
\centering
	\includegraphics[clip,trim = 0.2cm 0.2cm 0.2cm 0.2cm,width=.85\linewidth]{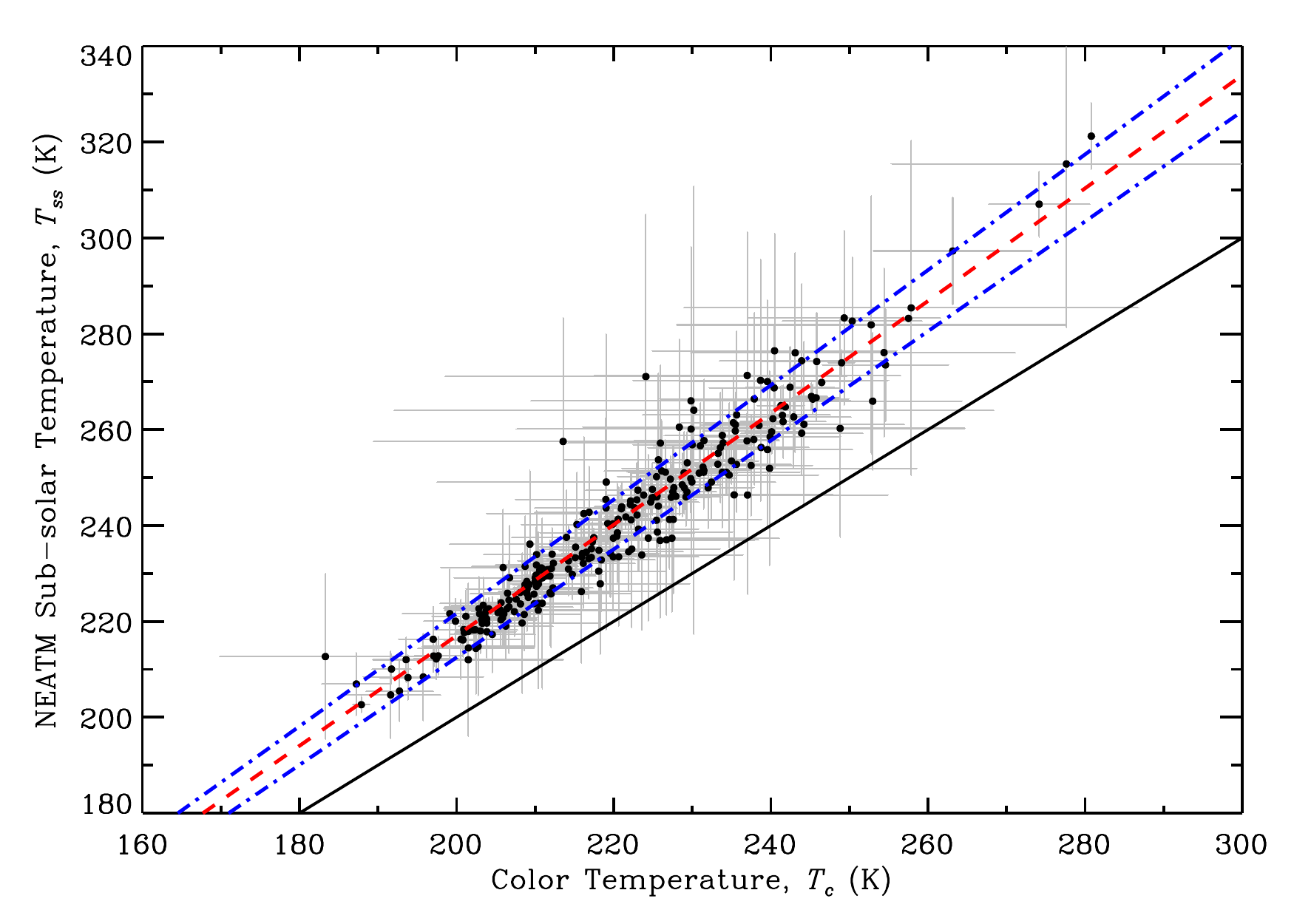}
	\caption{NEATM $T_{ss}$ as a function of color temperature, fit by the red dashed line (equation given in the text) with the blue dash-dot lines showing the 1$\sigma$ uncertainty in the fit parameters. The black solid line shows the identity function.}\label{figTcolorneatm}
\end{figure}

\begin{figure}
  \centering
    \includegraphics[width=.8\linewidth]{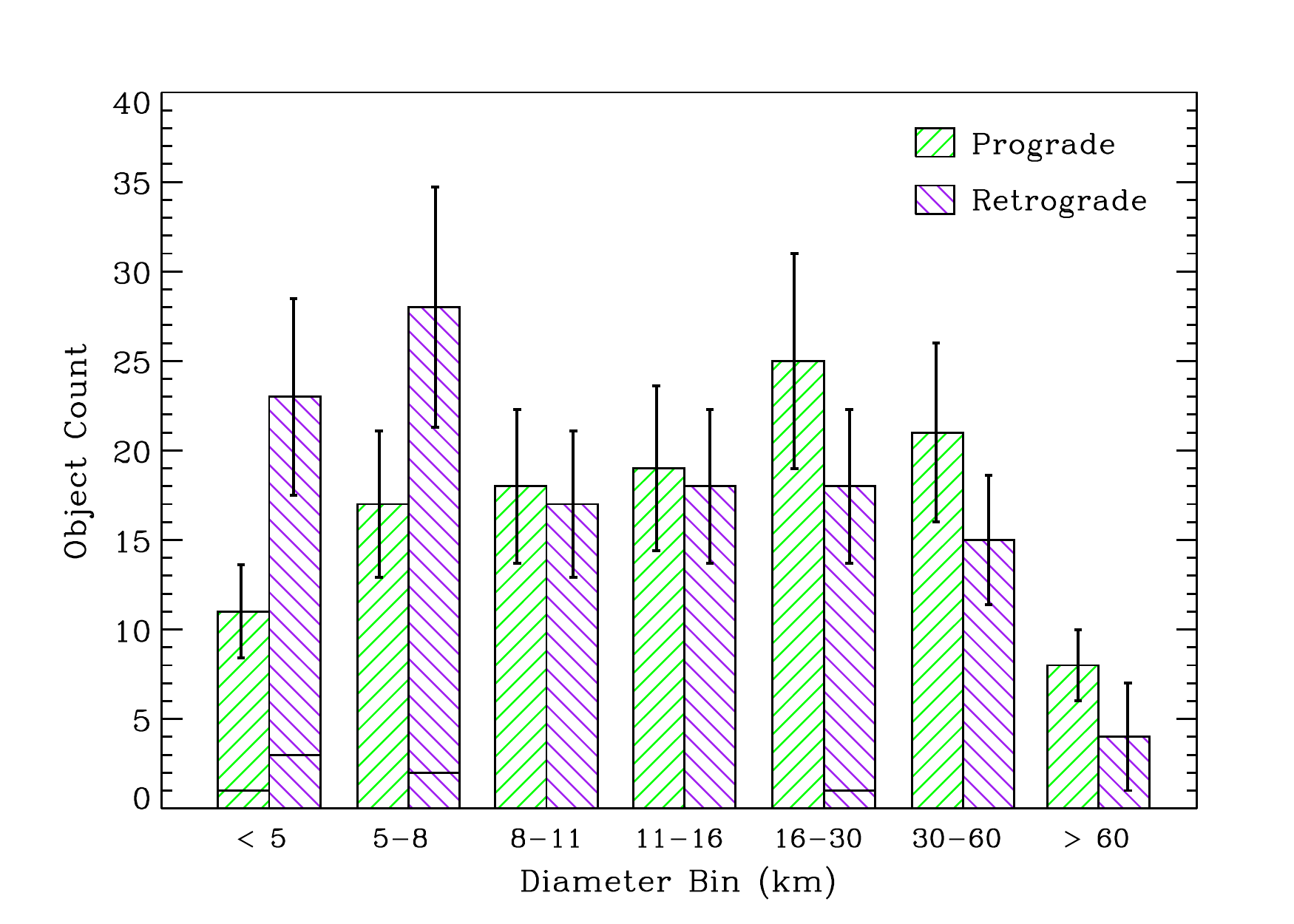}
  \caption{The number of prograde and retrograde rotators as a function of diameter bin. Horizontal lines indicate the number of NEAs within each bin. Vertical lines indicate the number of objects with indeterminate sense of spin within that size range, or one; whichever is greater.}\label{fig:proretro}
\end{figure}

\begin{figure}
  \centering
    \includegraphics[clip,trim = 0.1cm 0.1cm .8cm 1cm,width=.85\linewidth]{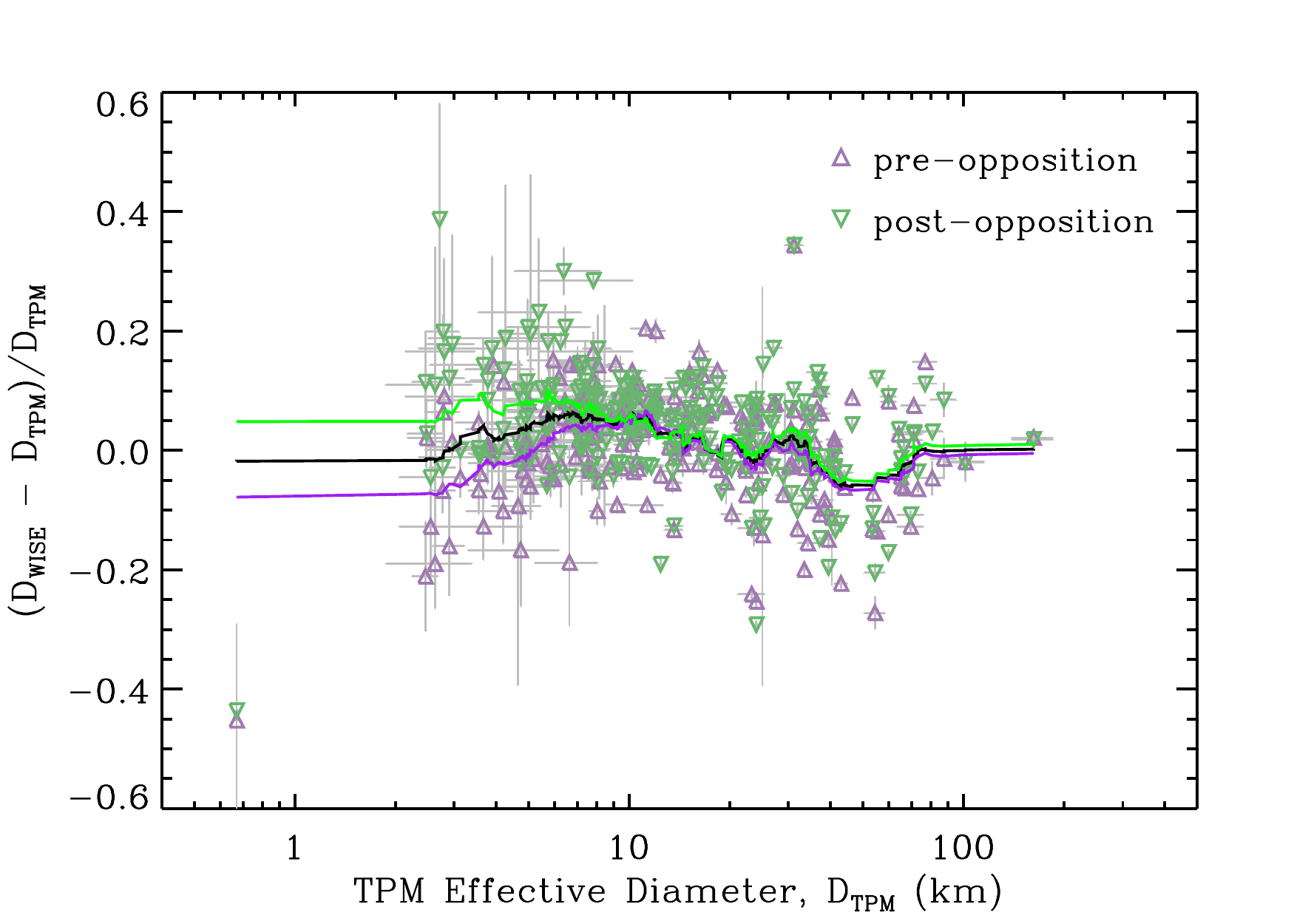}
  \caption[Comparison of TPM-derived diameters to the WISE team's NEATM fits.]{Comparison of the effective diameter values obtained by \cite{Masiero_etal11} and \cite{Mainzer_etal11b} to our reported TPM values. We plot the difference between the individual pre- and post-opposition diameters of the WISE team and our TPM diameter as a function of the diameter from our TPM. Purple, upward-facing and green, downward-facing triangles are data collected at pre- and post-opposition, respectively. the green line shows a running mean of the post-opposition data, the purple line shows a running mean of the pre-opposition data, and the black line shows a running mean of the relative diameter difference for all objects.}\label{fig:diamcomp2}
\end{figure}

\begin{figure}
  \centering
    \includegraphics[width=.8\linewidth]{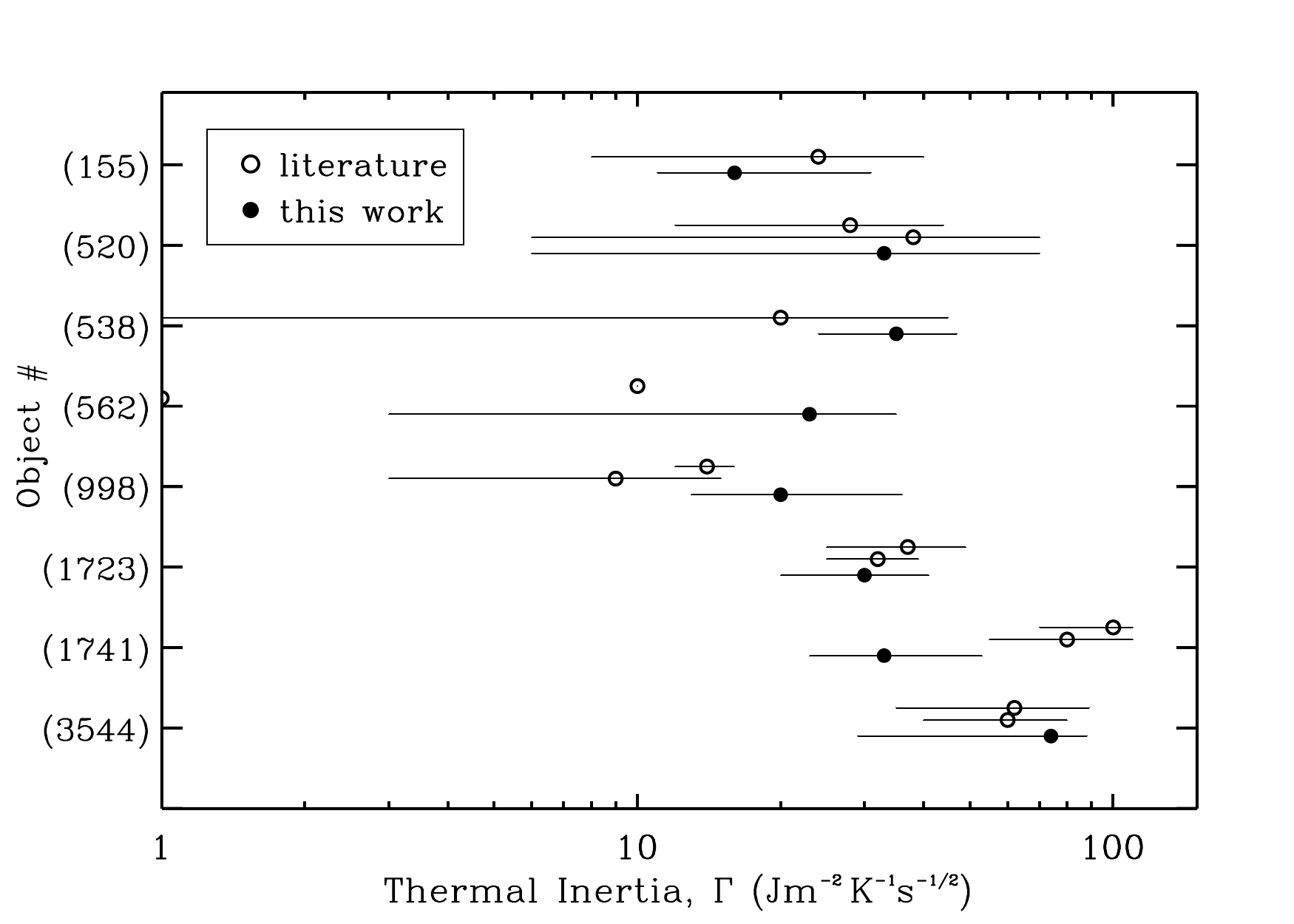}
  \caption{Comparison of literature thermal inertia values (open circles; \autoref{tab:prevTPM}) for individual objects to estimates in this work (filled circles; \autoref{tab:TPMresults}). All objects with previous estimates are from \citet{Hanus_etal15}, except (538) Fredicke \citep{Marciniak_etal19} and (1741) Giclas \citep{Pravec_etal19}.}\label{fig:TIcomp}
\end{figure}

\begin{landscape}

\begin{figure}
  \centering
	\includegraphics[clip,trim = 0.05cm 0cm 0.6cm 0.2cm,height=.3\linewidth]{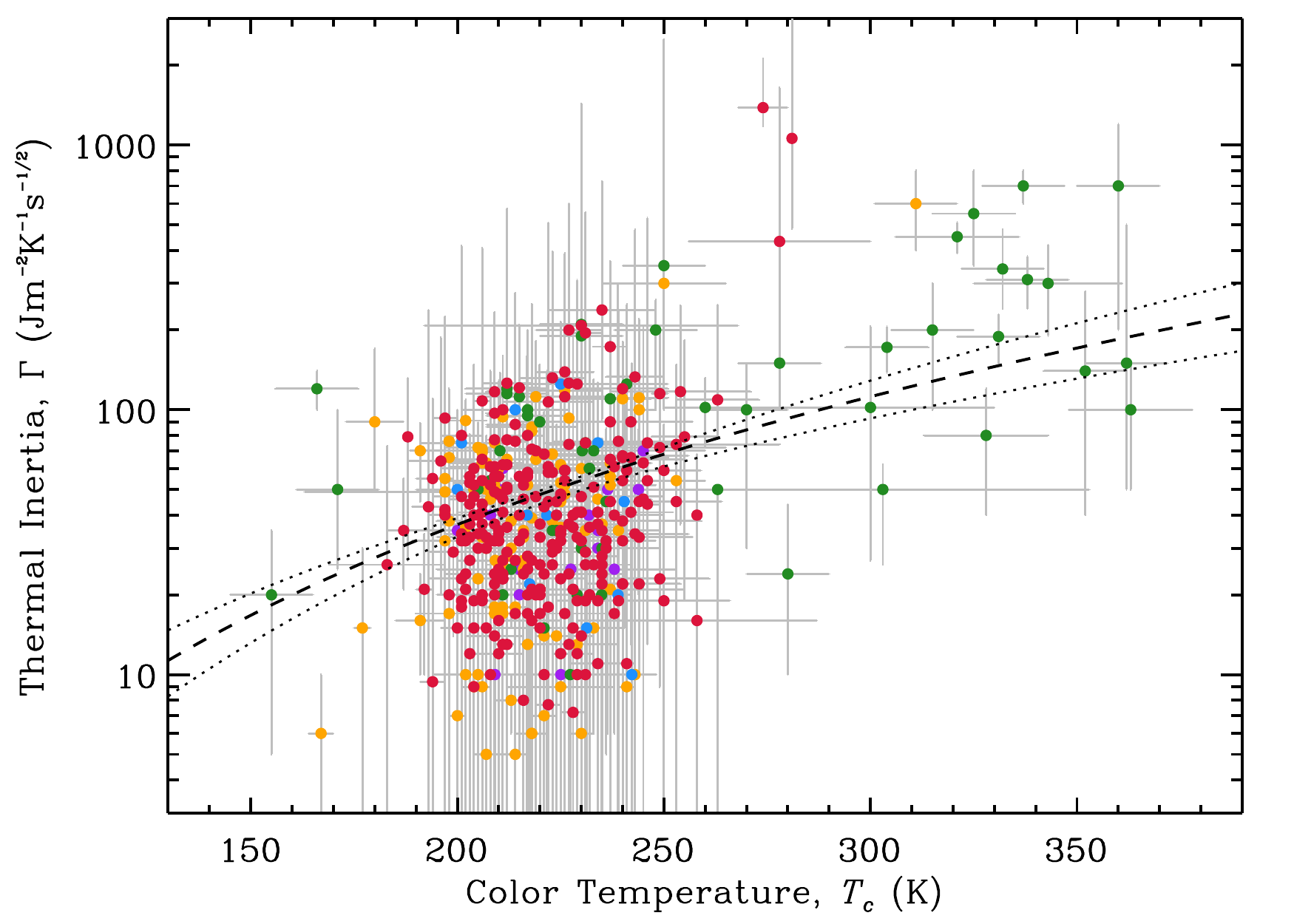}\includegraphics[clip,trim = 2.2cm 0cm 0.6cm 0.2cm, height=.3\linewidth]{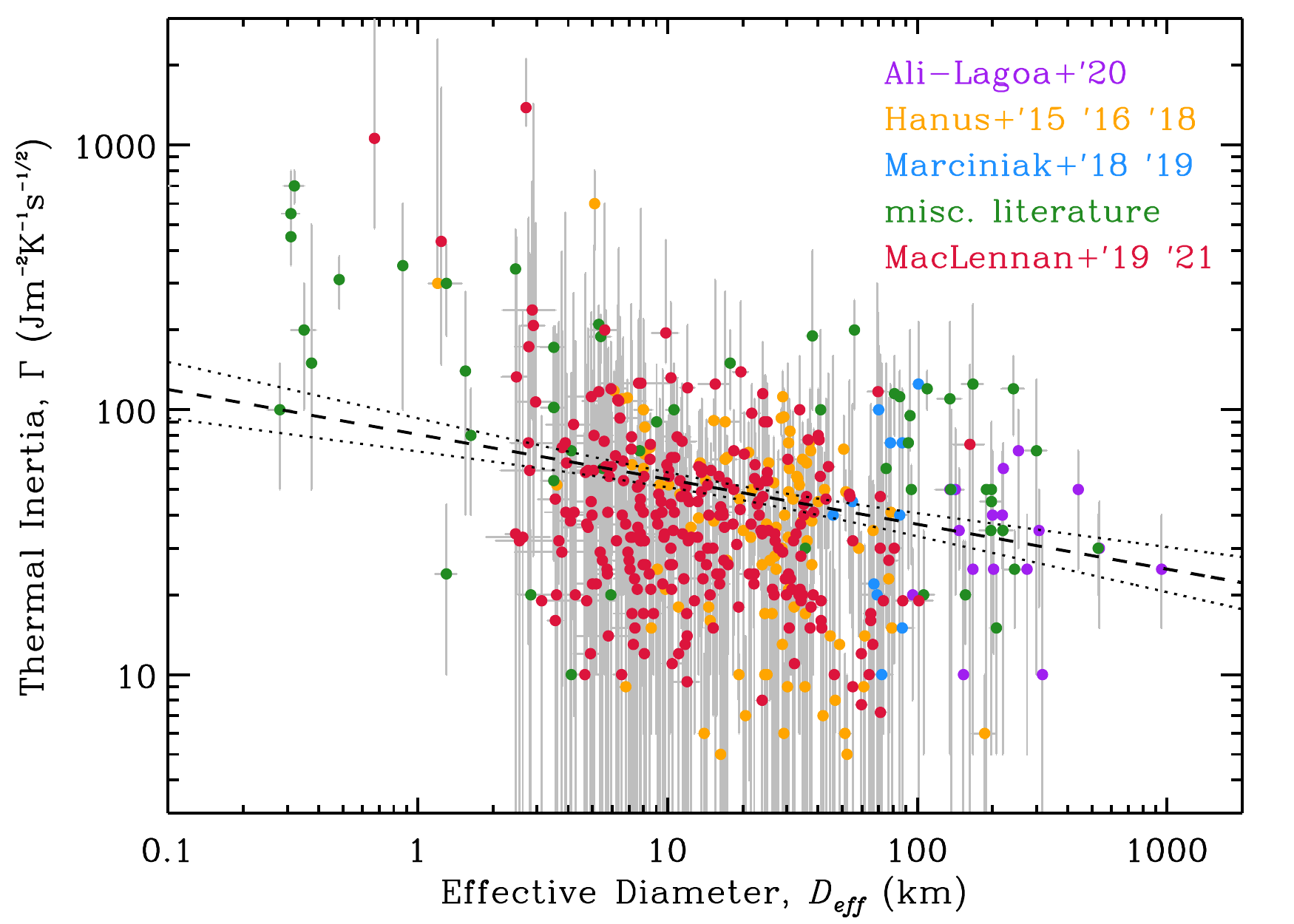}
	\includegraphics[clip,trim = 0.05cm 0cm 0.6cm 0.2cm,height=.3\linewidth]{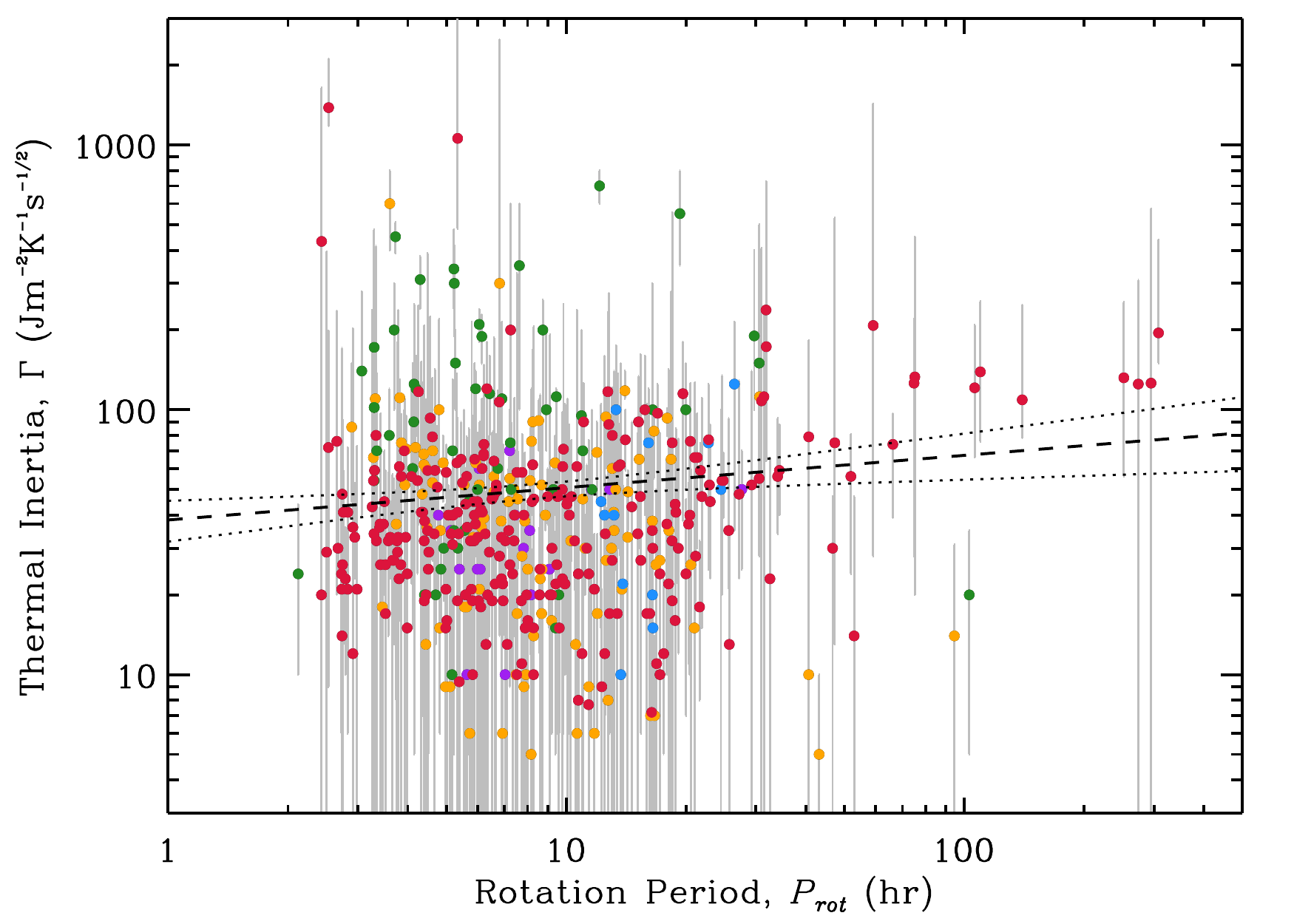}
  \caption{Thermal inertia dependence on surface (color) temperature, object diameter, and rotation period. The black lines show the best-fit multivariate regression model to the data (dashed) and 1$\sigma$ uncertainty (dotted). Red points are objects from this work (Maclennan'21; \autoref{tab:TPMresults}) and \citet{MacLennan+Emery19}. Other colors indicate the source from other works (\autoref{tab:prevTPM}), as indicated in the upper right panel, where misc. literature refers to papers that present less than 5 thermal inertias.}\label{fig:TImod}
\end{figure}

\end{landscape}

\begin{landscape}
\pagestyle{mylandscape}
\begin{scriptsize}
\begingroup
\setlength{\LTleft}{-50cm plus -1fill}
\setlength{\LTright}{\LTleft}


\endgroup
\end{footnotesize}

}







\end{document}